\documentclass[prb,showpacs,floatfix,twocolumn]{revtex4}
\usepackage{amsfonts}
\usepackage{stmaryrd}
\usepackage{bbm}
\usepackage{mathrsfs}
\usepackage{tipa}
\usepackage{amssymb}
\usepackage{txfonts}
\usepackage{graphicx}
\usepackage{dcolumn}
\usepackage{epstopdf}
\usepackage{array}

\newcommand{\PreserveBackslash}[1]{\let\temp=\\#1\let\\=\temp}
\newcolumntype{C}[1]{>{\PreserveBackslash\centering}p{#1}}
\newcolumntype{R}[1]{>{\PreserveBackslash\raggedleft}p{#1}}
\newcolumntype{L}[1]{>{\PreserveBackslash\raggedright}p{#1}}

\begin{document}

\newcommand*{\cm}{cm$^{-1}$\,}

%\preprint{APS/123-QED}

\title{Single crystal growth and optical conductivity of SrPt$_2$As$_2$ superconductors}
\author{A. F. Fang}
\author{T. Dong}
\author{H. P. Wang}
\author{Z. G. Chen}
\author{B. Cheng}
\author{Y. G. Shi}
\author{P. Zheng}
\author{G. Xu}
\author{L. Wang}
\author{J. Q. Li}
\author{N. L. Wang}

\affiliation{Beijing National Laboratory for Condensed Matter
Physics, Institute of Physics, Chinese Academy of Sciences,
Beijing 100190, People's Republic of China}
%


\begin{abstract}
SrPt$_2$As$_2$ single crystals with CaBe$_2$Ge$_2$-type structure
were synthesized by self-melting technique. X-ray diffraction,
transmission electron microscopy, electrical resistivity, magnetic susceptibility, specific
heat and optical spectroscopy measurements were conducted to
elucidate the properties of SrPt$_2$As$_2$. SrPt$_2$As$_2$ single
crystals undergo a structural phase transition well above the room
temperature (about 450 K) and become superconducting at 5.18 K.
The superconducting and structural phase
transition temperatures are reduced by 6$\%$ Iridium doping.
Both the pure SrPt$_2$As$_2$ and the Ir-doped single crystals are demonstrated to
be highly metallic with rather high plasma frequencies. In particular, the
optical spectroscopy measurement revealed two gap-like suppression structures:
a stronger one at high energy near 12000 \cm ($\sim$1.5 eV), and a less
prominent one at lower energy near 3200 \cm ($\sim$0.4 eV) for the pure compound.
We elaborate that the former is related to the correlation effect, while the
latter could be attributed to the partial energy gap formation
associated with structural phase transition.
\end{abstract}


\maketitle

\section{Introduction}
The discovery of high-temperature superconductivity in
iron-pnictide compounds has generated tremendous interest in the
condensed matter community.\cite{Hosono} Following the initial
discovery in LaOFeAs, a number of different structure types of
iron-based superconducting compounds that shared similar
properties were quickly discovered. Among the different systems,
the ThCr$_2$Si$_2$-type 122 pnictide compounds have attracted much
attention \cite{122-1,122-2,homes} due to their wide range of dopings
either by holes or electrons, as well as the easy growth of single
crystals with sufficient sizes for various experiments. The
peculiar properties of iron-pnictides come from the edge-sharing
Fe$_2$As$_2$ tetrahedral layers where Fe ion locates in the center
of each As tetrahedron (FeAs$_4$). Those tetrahedral layers are
separated by alkaline-earth elements Ca, Sr, and Ba or Eu in those
122 structures. In addition, to gain further insight into the
nature of the materials, many efforts were also made to explore
other 3d, 4d or 5d transitional metal based 122 compounds.

The 5d-transition metal Pt-pnictide
SrPt$_2$As$_2$ appears to be a very interesting material. It
shows superconductivity at 5.2 K.\cite{polymorph}
Different from other 122 superconducting compounds, SrPt$_2$As$_2$
does not crystallize in a ThCr$_2$Si$_2$ structure but adopt a
tetragonal CaBe$_2$Ge$_2$-type structure. The major difference
between ThCr$_2$Si$_2$ and CaBe$_2$Ge$_2$-type structures is that
the former contains two separate Fe$_2$As$_2$ tetrahedral layers
with Fe ion locating in the center of each As tetrahedron
(FeAs$_4$), while the latter contains only one Pt$_2$As$_2$ layer
with Pt ion locating in the center of each As tetrahedron
(PtAs$_4$) in an unit cell. Another Pt$_2$As$_2$ tetrahedral layer
is replaced by the As$_2$Pt$_2$ tetrahedral layer with As ion
locating in the center of each Pt tetrahedron (AsPt$_4$) (see the
structure in the inset of Fig. 1). Looking along the c-axis, the
AM$_2$As$_2$ with ThCr$_2$Si$_2$-type structure contains stacking
of a sequence of atomic sheets of
[As-M$_2$-As]-A-[As-M$_2$-As]..., while the SrPt$_2$As$_2$ with
CaBe$_2$Ge$_2$-type structure contains stacking of a sequence of
atomic sheets of [As-Pt$_2$-As]-Sr-[Pt-As$_2$-Pt]....

Another significant feature for the SrPt$_2$As$_2$ is that the
compound experiences a structural phase transition at about 470 K.
\cite{XRD} Below this temperature, a superstructure with a
modulation vector \textbf{q}=0.62\textbf{a}* is formed. Notably,
the modulation appears in the layers of the PtAs$_4$ tetrahedra,
the other atoms are only slightly affected by the modulation.
\cite{XRD} The structural phase transition was identified as a
charge-density-wave (CDW) instability in earlier work. As a
result, SrPt$_2$As$_2$ was argued to exhibit a coexistence of
superconductivity and CDW instability. \cite{polymorph}

It would be very interesting to explore the electronic properties
of SrPt$_2$As$_2$. Up to now there have been only a few
experimental studies on this compound. All of them were performed
on polycrystalline samples. In this work, we report on single
crystal growth of pure and 6$\%$ Iridium-doped SrPt$_2$As$_2$
compounds and investigation of their physical properties by
different experimental probes. We find that the structural phase
transition leads to a removal of about 17$\%$
spectral weight at low frequency in optical conductivity spectra, suggesting a
reconstruction of the Fermi surfaces.

\section{\label{sec:level2}Crystal growth and characterizations}

Single crystals of SrPt$_2$As$_2$ were synthesized by self-melting
technique. Sr (99.99\%) pieces, sponge Pt (99.99\%) grains and As
(99.99\%) grains were used as starting materials. PtAs precursor
was first fabricated by heating Pt grains and As powder at
700$^\circ \mathrm{C}$ sealed in a quartz tube. The fabricated
PtAs was grounded, mixed with Sr pieces in an atomic ratio of
2:1, and placed in an alumina crucible which was
sealed in an tantalum tube under Ar gas at the pressure of 1 bar.
The mixture was heated at 900$^\circ \mathrm{C}$
and then at 1330$^\circ \mathrm{C}$ for 15 hours respectively, and
then slowly cooled to 1030$^\circ \mathrm{C}$ at a rate of
3$^\circ \mathrm{C}$/$h$. We also tried to prepare SrPt$_2$As$_2$
polyctystalline first and melt it at 1330$^\circ \mathrm{C}$. Both
methods can yield big shiny plate-like single crystals. It is worth to
mention that the resultant single crystals are quite three
dimensional (3D) and hard to be cleaved. The observations are
quite different from iron pnictide compounds crystallized in
ThCr$_2$Si$_2$ structure, which are plate-like and very easy to
cleave from the melts. Those seem to indicate that the Pt-As
bonding strength between neighboring As$_2$Pt$_2$ and Pt$_2$As$_2$
tetrahedral layers may be comparable to the Pt-As bonding strength
within As$_2$Pt$_2$ or Pt$_2$As$_2$ tetrahedral layers, a result
being consistent with the first principle calculations by
Shein et al..\cite{calculation} We also grew single crystals of
Ir-doped SrPt$_2$As$_2$ by the same method. The obtained samples have similar
characteristics as SrPt$_2$As$_2$.

\begin{figure}[b]
\scalebox{0.52} {\includegraphics [bb=570 20 8cm 21.5cm]{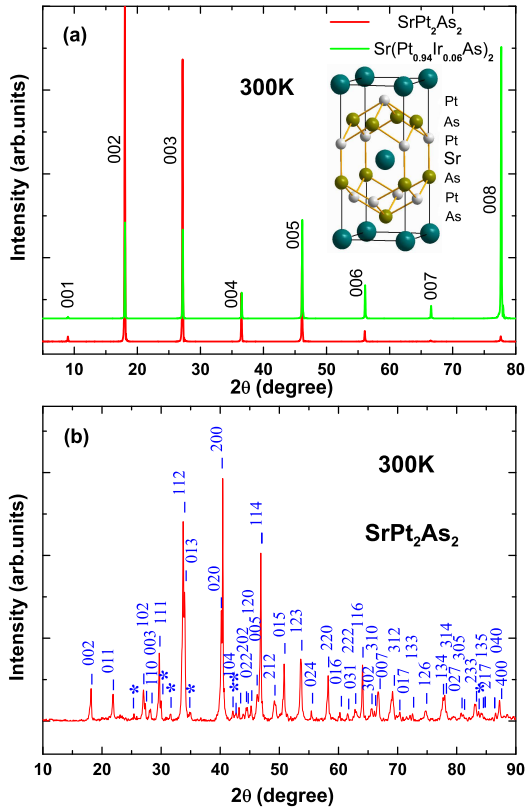}}
\caption{(Color online) (a) c-axis X-ray diffraction patterns of single
crystals SrPt$_2$As$_2$ and Sr(Pt$_{0.94}$Ir$_{0.06})_2$As$_2$
with crystal structure shown in the inset. (b) Powder XRD patterns of pulverized SrPt$_2$As$_2$ and indexing; the peaks marked by asterisks are from the contamination of impurity phases As and PtAs$_2$. }
\end{figure}

Figure 1 (a) shows the X-ray diffraction (XRD) patterns of SrPt$_2$As$_2$ and
Sr(Pt$_{0.94}$Ir$_{0.06})_2$As$_2$ single crystals at 300 K. The (00$l$)
($l$=$integer$) diffraction peaks indicate a good c-axis
orientation. Thus the largest face of the single crystals is the ab-plane.
The lattice parameters $c$$_1$=9.834$Å $ and
$c$$_2$=9.824$Å $ are obtained from the single crystal XRD data
of SrPt$_2$As$_2$ and Sr(Pt$_{0.94}$Ir$_{0.06})_2$As$_2$
respectively. The slight reduction in $c$-axis lattice parameter
comes possibly from the slight smaller ionic radius of element
Iridium. We also ground the crystal and performed power XRD
measurement. Figure (b) shows the XRD pattern of SrPt$_2$As$_2$ at 300 K. Since 300 K
is already below the structural phase transition, the superstructural modulation
would lead to a structural distortion from the high-temperature
tetragonal phase with the CaBe$_2$Ge$_2$-type structure to
a phase with the average structure of the
orthorhombic space group Pmmn.\cite{XRD, polymorph} The powder XRD
pattern of Fig. 1 (b) could be indexed by such averaged structure with
lattice parameters $a$=4.47$Å $, $b$=4.50$Å $ and $c$=9.82$Å $, being consistent with
the previous report.\cite{polymorph} The minor extra peaks labeled by asterisks
were from the contamination of impurity phases As and PtAs$_2$.
The structural characterizations reveal good quality of
the obtained single crystals.

\begin{figure}[t]
\scalebox{0.45} {\includegraphics [bb=580 15 8cm 20.5cm]{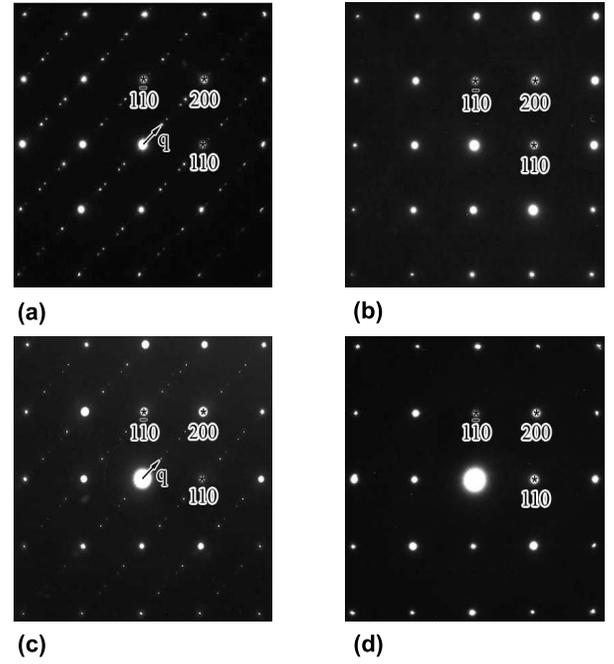}}
\caption{Electron diffraction patterns taken along
[001] zone-axis directions of SrPt$_2$As$_2$ and
Sr(Pt$_{0.94}$Ir$_{0.06})_2$As$_2$; (a) and (b) are for
SrPt$_2$As$_2$ at room temperature and 470 K, respectively,
conforming additional spots with a wave vector of q=0.62 a$^*$;
(c) and (d) the electron diffraction patterns
of Sr(Pt$_{0.94}$Ir$_{0.06})_2$As$_2$ at room temperature and 400 K, respectively. }
\end{figure}

The electron diffraction patterns were taken on a FEI Tecnai-F20
(200 kV) transmission electron microscope (TEM). The TEM samples
were prepared by crushing the single crystals, and then the
resultant suspensions were dispersed on a holey copper grid coated
with a thin carbon film. Figure 2 displays the room-temperature
electron diffraction patterns taken along [001] zone axis of
SrPt$_2$As$_2$ ((a) and (b)) and
Sr(Pt$_{0.94}$Ir$_{0.06})_2$As$_2$ ((c) and (d)), respectively.
The most striking structural phenomenon in this pattern is the
appearance of a series of satellite spots aligned with the main
diffraction spots, as clearly illustrated in Fig. (a) and (c).
These satellite spots in general can be characterized by the
modulation wave vector \textbf{q}$\approx$0.62\textbf{a}* for
SrPt$_2$As$_2$, as reported by Imre et. al.,\cite {XRD}. A slight reduction
of the modulation wave vector (roughly \textbf{q}$\approx$0.60\textbf{a}*) is observed for
Sr(Pt$_{0.94}$Ir$_{0.06})_2$As$_2$. The satellite spots were not
visible when measurements were performed at 470 K and 400 K on the
two samples, respectively.

\begin{figure}
\scalebox{0.64} {\includegraphics [bb=650 15 8cm
21.5cm]{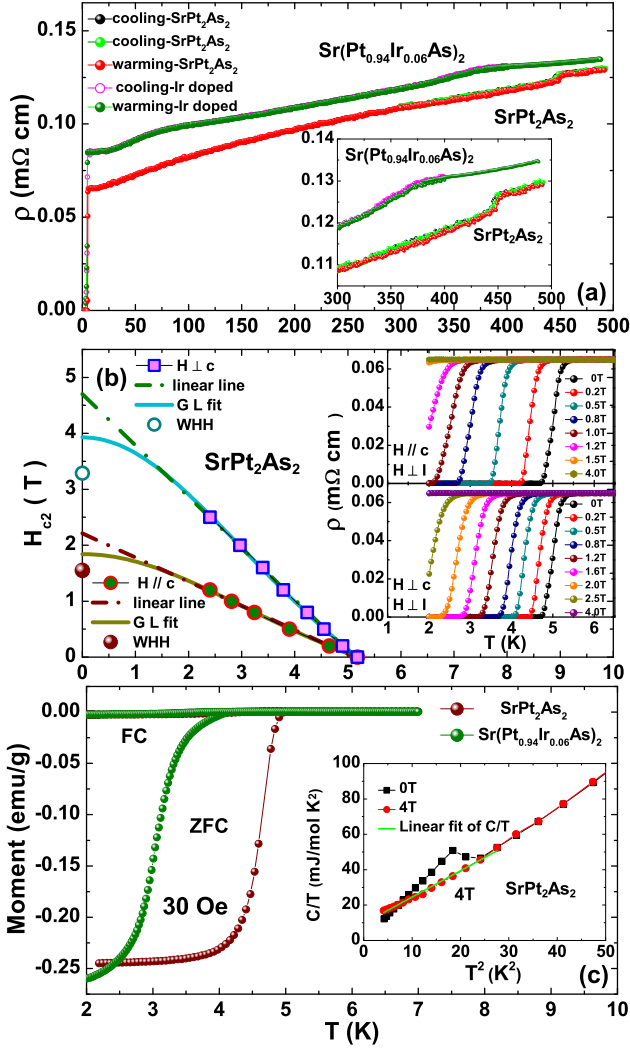}} \caption{(Color online) Transport data of
SrPt$_2$As$_2$ and Sr(Pt$_{0.94}$Ir$_{0.06})_2$As$_2$ vs
temperature. (a) Resistivity $\rho$ vs temperature for both
samples. Inset: the expanded region between 300 K and 500 K; (b) the temperature dependence of the upper
critical field $H_c{_2 }$. Inset shows the low temperature
resistivity $\rho$ under magnetic field up to 4T for
SrPt$_2$As$_2$; (c) the low temperature magnetization
and specific heat data for SrPt$_2$As$_2$. }
\end{figure}

Resistivity measurements were performed in both Quantum Design
physical properties measurement system (PPMS) (below 350 K) and a
home-made transport measurement system (up to 500 K) by a standard
dc four-probe method. Figure 3 presents the temperature dependence
of resistivity for SrPt$_2$As$_2$ and
Sr(Pt$_{0.94}$Ir$_{0.06})_2$As$_2$. The two samples show sharp
superconducting transitions at 5.18K and 4.73K (90$\%$ of the
normal state resistivity), respectively. At high temperatures, a
small resistivity jump near 450 K is seen for pure SrPt$_2$As$_2$
crystal with almost no hysteresis upon warming
and cooling processes, while only a slope change could be seen
near 375 K for Sr(Pt$_{0.94}$Ir$_{0.06})_2$As$_2$ sample. The high
temperature anomalies are of the characteristic of the first order
phase transition.

The insets of Fig. 3 (b) shows the temperature-dependent
resistivity curves in magnetic fields up to 4 T with $H//c$ and $H\bot c$, respectively. The superconducting transition shifts toward lower
temperature with an increase of the magnetic field, indicating a
field-induced pair breaking effect. Taking the criterion of 90\%
of $\rho_n$, the upper critical field $H_c{_2 }(T)$ can be
obtained as displayed in the middle panel of Fig. 3. According
to the Werthaner-Helfand-Hohenberg (WHH) formula,\cite{WHH} the
zero-temperature upper critical field $H_c{_2 }(0)$ can be
estimated by
\begin{equation}
{H_{c2}}(0) =  - 0.693{T_c}{\left(
{\frac{{d{H_{c2}}}}{{dT}}}\right)_{T = {T_C}}} . \label{chik}
\end{equation}
Taking $T_c$ = 5.18 K, ($dH_{c2}^{//c}/dT)_{T = {T_C}}\approx-0.43$ T/K and ($dH_{c2}^{\bot c}/dT)_{T = {T_C}}\approx-0.92$ T/K, $H_{c2}^{//c}(0)\approx1.55$ T, $H_{c2}^{\bot c}(0)\approx3.29$ T, and $(H_{c2} ^{\bot c}(0)/H_{c2}^{//c}(0))_{WHH}\approx2.12$ can be achieved.
Different from the quite high upper critical field  of the
FeAs-based superconductors,\cite{Hc2-1,Hc2-2} this PtAs-based
superconductor is quite sensitive to the magnetic field. The anisotropic ratio is
comparable to that of BaNi$_2$P$_2$ \cite{BaNi2P2}. The ratio
$\mu_0H_c{_2 }(0)/k_BT_c$=3.29/5.18 T/K$\approx0.66$ T/K is far
less than the Pauli limit $\mu_0H_c{_2 }(0)/k_BT_c=1.84$ T/K,
representing the singlet pairing if assuming a weak spin-orbital
coupling.\cite{Pauli limit} Therefore SrPt$_2$As$_2$ may be the
conventional superconductor in accord with the estimated result of
the average electron-phonon coupling constant.\cite{calculation}
We also determine $H_c{_2 }(T)$ using the formula based on the
Ginzburg-Landau(GL) equation, $H_{c2}(T) = H_{c2}(0)(1 -
{t^2})/(1 + {t^2})$. The obtaining values are $H_{c2}^{//c}(0)\approx1.84$ T, $H_{c2}^{\bot c}(0)\approx3.93$T and $(H_{c2}^{\bot c}(0)/H_{c2}^{//c}(0))_{GL}\approx2.14$,
which are close to the value obtained by WHH analysis.

The superconducting transition could also be seen clearly in the magnetization
and specific heat measurements, yielding evidence for bulk superconductivity,
as presented in Fig. 3(c). The magnetic susceptibility was measured in a Quantum
Design superconducting quantum interference device
vibrating-sample magnetometer system (SQUID-VSM). At the
temperature where resistivity drops to zero, the magnetic
susceptibility displays a sharp superconducting transition.
The specific-heat coefficient $C/T$
vs ${T^{^2}}$ relation for SrPt$_2$As$_2$ is shown in the inset of Fig. 3(c).
In agreement with the measurement on polycrystalline sample,\cite{polymorph}
the superconductivity is completely
suppressed by a magnetic field of 4 T. From the linear fit of $C=\gamma T+\beta T^3$ relation
at low temperature, we get the electronic specific heat coefficient
$\gamma=9.63(8)$ $mJ K^{-2}mol^{-1}$ and $\beta=1.48(3)$ $mJ K^{-4}mol^{-1}$.
The extracted Debye temperature is $\Theta=187 K$.

\begin{figure}
\includegraphics[clip,width=3.6in]{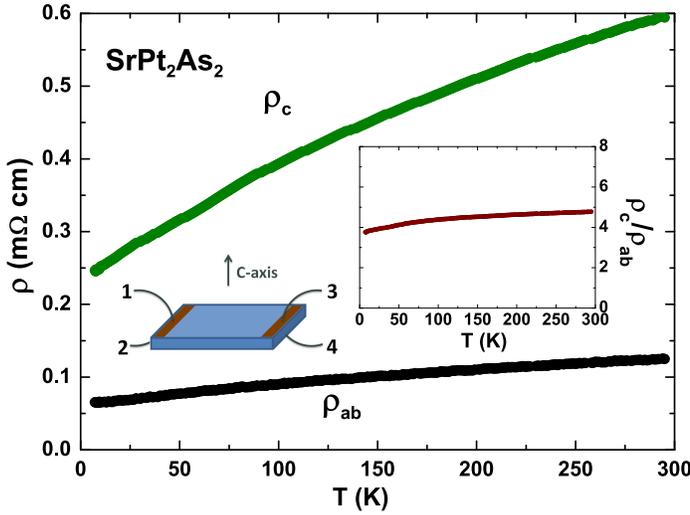}
\caption{(Color online)
Anisotropic resistivity vs temperature for SrPt$_2$As$_2$. Inset shows the
ratio of $\rho_c/\rho_a{_b}$. The contact configuration of
leads is also displayed.}
\end{figure}

We also measured the anisotropic resistivity between ab-plane and
the c-axis by a generalized Montgomery method,\cite{Montgomery} as presented in Fig. 4.
The contact configuration is shown in the inset. Instead of
the point contact in the original Montgomery method, we extended
the contacts along the edges of the sample to minimize the
distortion effect of finite contact size along the c axis
.\cite{wang-Resistance} The resistances, R$_1=V_{1,3}/I_{2,4}$ and R$_2=V_{1,2}/I_{3,4}$, with the
current nominally parallel and normal to the ab plane,
respectively, were measured by the four leads method. The
resistivity components, $\rho_{ab}$(T) and $\rho_c$(T), were then
calculated from the ratio of R$_2$/R$_1$ in the thin sample limit
of the Montgomery technique. The obtained $\rho_a{_b}$, $\rho_c$
and $\rho_c/\rho_a{_b}$ are shown in the main panel and
its inset, respectively. The anisotropic ratio $\rho_c/\rho_a{_b}$ is about 4.8
at 300 K. There is only a weak temperature dependence for the
anisotropic resistivity. The anisotropic ratio is reduced to 3.8
at the lowest measurement temperature, implying a slight
enhancement of three dimensionality. The values of $(\rho_{//c}/\rho_{\bot c})^{1/2}\sim$2.19 (300 K), 1.95 (low temperature) are quite close to the values of $H_c{_2 }^{\bot c}(0)/H_c{_2 }^{//c}(0)$ $(=(m_{//c}^*/m_{\bot c}^*)^{1/2})$.\cite{BaNi2P2}

\section{\label{sec:level2}Optical conductivity}

The ab-plane optical reflectance measurements were conducted on Bruker 113v
and Vertex 80v spectrometers in the frequency range from 40 to
40000 \cm. An \textit{in situ} gold and aluminum over-coating
technique was used to get the reflectivity R($\omega$). The real
part of conductivity $\sigma_1(\omega)$ was obtained by the
Kramers-Kronig transformation of R($\omega$). The Hagen-Rubens
relation was used for low frequency extrapolation; at high
frequency side a $\omega^{-1}$ relation was used up to 300000 \cm,
above which $\omega^{-4}$ was applied.

Figure 5 illustrates the reflectance curves below 20000 \cm for
several temperatures from 10 to 500 K, while the inset displays
the enlarged spectral behavior up to 5000 \cm. Being consistent with
the dc resistivity data, the optical reflectance shows good metallic response
in both frequency and temperature dependence. The overall values of R($\omega$) are much higher
than that of iron-pnictides, e.g. the prototype AFe$_2$As$_2$ (A=Ba, Sr),\cite{BaFe2As2}
and are also higher than that of BaNi$_2$As$_2$.\cite{BaNi2As2} Furthermore,
the reflectance minimum (or overdamped edge) extends to much higher frequency. Those characteristics
indicate a rather high plasma frequency. An estimation will be made below.

\begin{figure}[b]
\scalebox{0.4} {\includegraphics [bb=600 20 8cm
21.5cm]{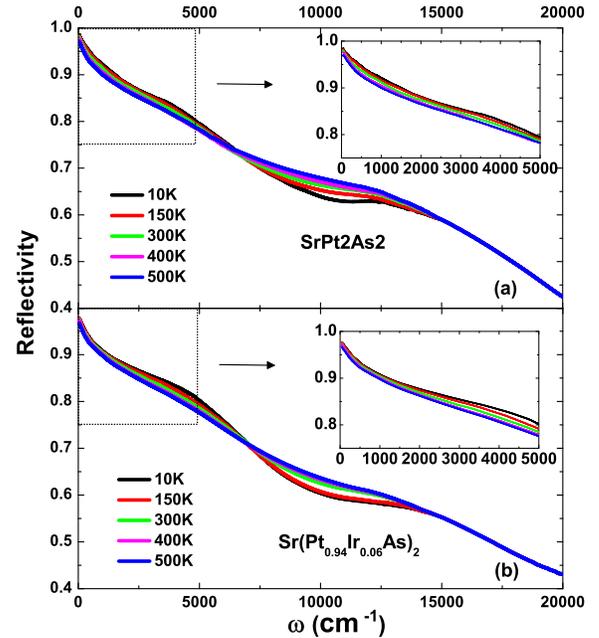}} \caption{(Color online) The reflectance
spectra of single crystals (a) SrPt$_2$As$_2$ and (b) Sr(Pt$_{0.94}$Ir$_{0.06})_2$As$_2$
 below 20000 \cm; Inset: R($\omega$) up to 5000 \cm.}
\end{figure}

The most interesting observation in the R($\omega$)
is the manifestation of the two reverse S-like
suppression features at the reduced temperatures. A stronger suppression
structure appears at high energy near 12000 \cm ($\sim$1.5 eV), and a less prominent suppression
appears at lower energy scale near 3200 \cm ($\sim$0.4 eV) for SrPt$_2$As$_2$.
For Ir doped compound Sr(Pt$_{0.94}$Ir$_{0.06})_2$As$_2$, the latter feature
shifts to lower frequency. Understanding the two features
is of crucial importance for understanding the evolution of the electronic structure across
the transition.

\begin{figure*}
\scalebox{0.6} {\includegraphics [bb=625 30 8cm
17.5cm]{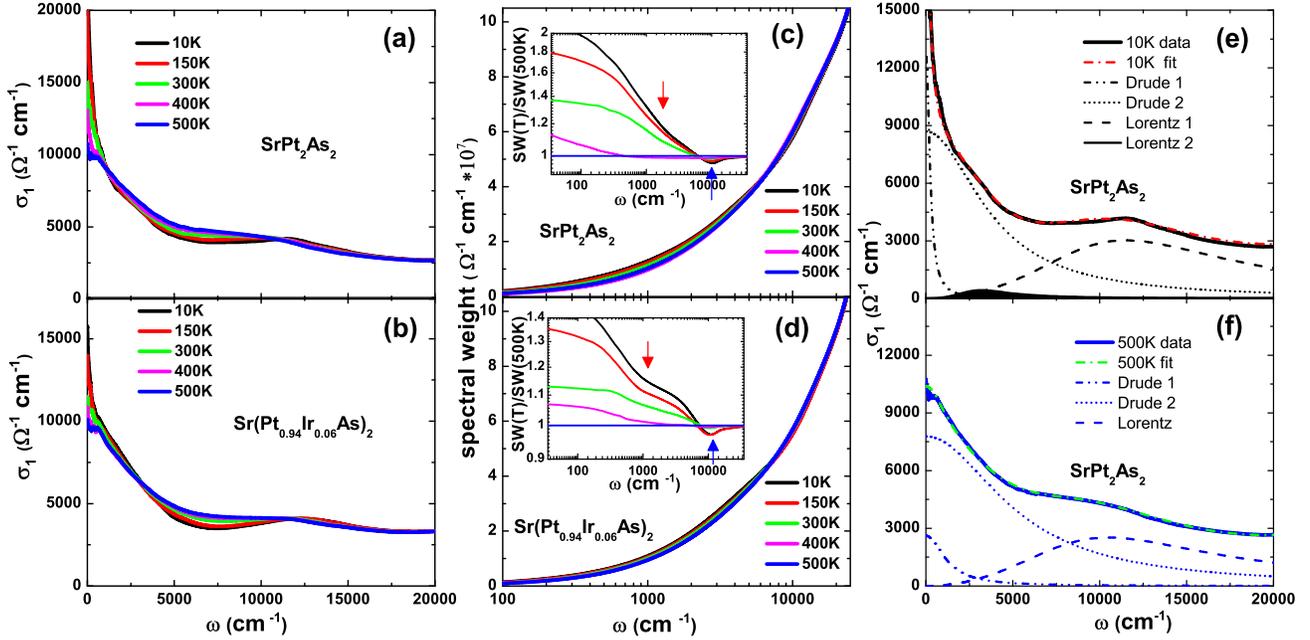}} \caption{(Color online) Left panel: the
temperature dependence of the real part of the optical
conductivity $\sigma_1(\omega)$ for (a)SrPt$_2$As$_2$ and
(b)Sr(Pt$_{0.94}$Ir$_{0.06})_2$As$_2$ up to 20000 \cm. Middle
panel: The temperature dependence of the integrated spectral
weight up to 25000 \cm for (c)SrPt$_2$As$_2$ and
(d)Sr(Pt$_{0.94}$Ir$_{0.06})_2$As$_2$. Inset: the normalized
spectral weight SW(T)/SW(500K). Right panel: the experimental data
of $\sigma_1(\omega)$ and the Drude-Lorentz fit at 10 K (e) and 500 K (f)
for SrPt$_2$As$_2$.}
\end{figure*}

The two suppression features also manifest in the optical
conductivity spectrum $\sigma_1(\omega)$,
as presented in Fig. 6 (a) and (b). The Drude-like conductivity can be observed for all
spectra at low frequencies. Corresponding to the high energy suppression structure
of R($\omega$), the conductivity spectra display a broad peak near 12000 \cm. The
suppressed spectral weight below this energy is transferred to
higher energy scales. By lowering the temperature, a weak bump
related to the low frequency suppression of the
reflectance could be also seen in $\sigma_1(\omega)$. In order to have a better
view about the evolution of the optical conductivity, we plot the
integrated spectral weight distribution in Fig. (c) and (d) in a semilog scale. The insets present the
normalized spectral weight SW(T)/SW(500K). Due to the metallic
response the spectral weight gradually increases with decreasing
the temperature in very low frequency range. However, a pronounced drop
appears in the spectral weight curves near 2000 \cm followed by a
shoulder-like feature, as indicated by the red
arrow. The more prominent structure of the high energy suppression in
$\sigma_1(\omega)$ at high energy scale leads to a dip feature in
normalized spectral weight as indicated by the blue arrow.

It is well known that the interband transitions contribute
dominantly to the conductivity spectrum at high energies and usually lead peaks
in $\sigma_1(\omega)$. However, one would not expect to see a prominent
temperature dependence for an interband transition. So the temperature dependent
spectral weight transfer near 1.5 eV must have a different origin. In
general, the temperature-induced reverse S-like suppression in R($\omega$)
and corresponding peak-like feature in $\sigma_1(\omega)$ would represent the
formation of an energy gap (or pseudogap). Nevertheless, the energy scale of the
high energy gap feature near 1.5 eV is too high to be connected to the structural
phase transition at 450 K or 375 K for the two compounds. A careful inspection
of the $\sigma_1(\omega)$ spectra indicates that
the characteristic peak-like lineshape is already present at 500 K.
It deserves to remark that similar spectral
suppression features are also present in many other
correlated electronic systems, for example, the Fe-pnictide compounds
though the corresponding energy scale is much lower,
roughly at 0.6 eV.\cite{nlwang,Schafgans} In general, we could attribute the
temperature-dependent high energy feature to the correlation effect, i.e.,
the quasi-particles contain not only the
coherent spectral weight at low energy but also the incoherent
part at high energies arising from the strong Coulomb
interactions.\cite{Georges,Rozenberg,Qimiao} For Fe-pnictides, because of the
relatively narrow 3d electron bands, the Hund's coupling effect between
different orbitals plays a major role.\cite{nlwang,Schafgans} For SrPt$_2$As$_2$ compound,
the 5d orbitals are spatially much extended than 3d or 4d compounds,
the Hund's rule coupling interaction is relatively weak, we speculate that
the peak feature at 1.5 eV is mainly caused by the
on-site Coulomb repulsion (Hubbard U) effect.\cite{Georges,Rozenberg}

\begin{table}[htbp]
\begin{center}
\newsavebox{\tablebox}
\begin{lrbox}{\tablebox}
\begin{tabular}{*{12}{C{13mm}}}
%\arrayrulewidth=1pt
\hline \hline\ \\[-2ex]
SrPt$_2$As$_2$&$\omega_p{_1}$&$\gamma_D{_1}$&$\omega_p{_2}$&$\gamma_D{_1}$&$\omega_p$\\[4pt]
\hline
\\[-3ex]
10K&$1.40\times10^4$&260&$4.45\times10^4$&3800&$4.67\times10^4$\\[4pt]
150K&$1.40\times10^4$&360&$4.46\times10^4$&3900&$4.67\times10^4$\\[4pt]
300K&$1.40\times10^4$&570&$4.48\times10^4$&4000&$4.69\times10^4$\\[4pt]
400K&$1.40\times10^4$&700&$4.50\times10^4$&4200&$4.71\times10^4$\\[4pt]
500K&$1.40\times10^4$&1250&$4.95\times10^4$&5250&$5.14\times10^4$\\[4pt]
\hline \hline
\end{tabular}
\end{lrbox}
\caption{The fitting parameters of the two Drude components for
SrPt$_2$As$_2$ at selected
temperatures. The units for $\omega_p{_ i}$ and $\gamma_D{_ i}$ is
\cm.} \scalebox{1.0}{\usebox{\tablebox}}
\end{center}
\end{table}

On the other hand, the spectral weight suppression feature at
lower energy scale near 3200 \cm is likely to be caused by the removal of
a small part of Fermi surface below the structural transition. It is noted that, unlike
other compounds showing first order structural phase transitions,
e.g. BaNi$_2$As$_2$ \cite{BaNi2As2} or IrTe$_2$ \cite{Fang},
where the optical conductivity spectra show sudden and dramatic changes over broad frequency across
the phase transitions due to the reconstruction
of the band structures, the spectral suppression feature in the present case is
rather weak and evolves continuously with
temperature, being more similar to some CDW materials
such as 2H-TaS$_2$ and 2H-NbSe$_2$.\cite{NbSe2-Basov, NbSe2-Hu, TaS2-Hu}

To quantify the spectral change, particularly the evolution of the Drude
component, across the phase transition, we tried to decompose the optical conductivity
spectral into different components using a Drude-Lorentz analysis.
The dielectric function has the form
\begin{equation}
\epsilon(\omega)=\epsilon_\infty-\sum_{i}{{\omega_{p,i}^2}\over{\omega_i^2+i\omega/\tau_i}}+\sum_{j}{{\Omega_j^2}\over{\omega_j^2-\omega^2-i\omega/\tau_j}}.
\label{chik}
\end{equation}
where $\epsilon_\infty$ is the real part of dielectric constant at
high energy, $\omega_{p,i}$ and $1/\tau_i$ are the plasma frequency
and scattering rate of the itinerant carriers in the $i$th band
respectively, and $\omega_j$, $\Omega_j$ and $1/\tau_j$ are the
resonance, strength and width of the $j$th Lorentz oscillator. The
Drude components represent the contribution from conduction
electrons, while the Lorentz components describe the interband
transitions. We found that the optical conductivity spectrum at 500 K below
15000 \cm could be reasonably reproduced by two Drude and one
Lorentz components. However, at low temperatures, an additional
Lorentz component centered at 3200 \cm should be better added.
Fig. 6 (e) and (f) shows the conductivity spectra at 10 K
and 500 K together with the Drude-Lorentz fitting components for SrPt$_2$As$_2$.
The parameters of Drude components at different temperatures are listed in
Table I. We find that the two Drude components show
ordinary narrowing with decreasing temperature due to the metallic response, while the
gapping of the Fermi surfaces mainly appears in the broad Drude component.
We can use the formula $\omega_p=\sqrt{{\omega_p{_1}^2} + {\omega_p{_2}^2}}$ to estimate the
overall plasma frequency, then we get $\omega _P\approx$ 51400 cm$^{-1}$ at 500 K, and
46650 cm$^{-1}$ at 10 K, respectively. Those values are much higher than that of
Fe-pnictide superconductors. The reduction of the overall plasma frequency could be
attributed to the gapping of the Fermi surface. Roughly, we can
use ($\omega _P^2(500K)-\omega _P^2(10K))/\omega
_P^2(500K)\approx$17\% to
estimate the missing carrier density. The missing spectral weight is mainly
transferred to the Lorentz part centered at 3200 \cm.
It is worth to mention that our decomposition of conductivity spectrum into
two Drude components may lead to an over-estimate of the spectral weights (particularly the
broad one) or the plasma frequencies, but this effect is minor.
Even if we try a single Drude component to reproduce the low-frequency spectral weight,
we still obtain a plasma frequency close to 50000 cm$^{-1}$ at 500 K. Nevertheless, the one
Drude component could not well reproduce the conductivity spectra, in particular those at
low temperatures. Another source
of uncertainty in the estimation of the plasma frequency is
from the Kramers-Kronig transformation of the reflectance
spectra. Different high frequency extrapolations could affect
the spectral weight even at low frequencies, thus resulting in
somewhat different values of the plasma frequencies.

At present, it is not clear whether the compounds have nested Fermi surfaces;
and if nesting exists, whether the nesting wave vector matches with the structural modulation
wave vector. In other words, whether or not the nesting instability is strong enough to
drive the structural distortion. Further theoretical and experimental studies on this system
are needed.

\section{\label{sec:level2}CONCLUSIONS}

We have successfully grown single crystals of SrPt$_2$As$_2$ using a
self-melting technique and conducted careful characterizations by X-ray diffraction,
transmission electron microscopy, electrical resistivity, magnetic susceptibility and specific
heat measurements. SrPt$_2$As$_2$ single crystals are manifested to possess a bulk superconductivity and carry a superstructure with a modulation vector \textbf{q}=0.62\textbf{a}* below 5.18K and 450K, respectively. The optical spectroscopy study on the pure and Ir-doped SrPt$_2$As$_2$
revealed two gap-like suppression structures.
The one at higher energy scale is attributed to the correlation effect, while the
other one at lower energy scale is suggested to be the partial energy gap formation
associated with structural phase transition. It is estimated that roughly
17\% carrier density was removed accompanying with the energy gap formation.

\begin{center}
\small{\textbf{ACKNOWLEDGMENTS}}
\end{center}
This work was supported by the National Science Foundation
of China (10834013, 11074291, 11120101003) and the 973 project
of the Ministry of Science and Technology of China
(2011CB921701).

\end{document}